\documentclass[pra,amsmath,amssymb,twocolumn]{revtex4}
\usepackage{amsmath}
\usepackage{amssymb}

\usepackage{graphicx}
\input epsf.tex
\usepackage{amsthm}

\newcommand{\be}{\begin{equation}}
\newcommand{\ee}{\end{equation}}
\newcommand{\ba}{\begin{eqnarray}}
\newcommand{\ea}{\end{eqnarray}}
\newcommand{\ban}{\begin{eqnarray*}}
\newcommand{\ean}{\end{eqnarray*}}

\newcommand{\one}{\leavevmode\hbox{\small1\normalsize\kern-.33em1}}

\begin{document}

\title{The black paper of quantum cryptography: real implementation problems}

\author{Valerio Scarani, Christian Kurtsiefer}
\affiliation{Centre for Quantum Technologies \& Department of Physics,
National University of Singapore, Singapore}

\date{\today}

\begin{abstract}
The laws of physics play a crucial role in the security of quantum key distribution (QKD). This fact has often been misunderstood as if the security of QKD would be based \textit{only} on the laws of physics. As the experts know well, things are more subtle. We review the progresses in practical QKD focusing on (I) the elements of trust that are common to classical and quantum implementations of key distribution; and (II) some threats to security that have been highlighted recently, none of which is unredeemable (i.e., in principle QKD can be made secure). This leads us to guess that the field, similar to non-quantum modern cryptography, is going to split in two directions: those who pursue practical devices may have to moderate their security claims; those who pursue ultimate security may have to suspend their claims of usefulness.
\end{abstract}
\maketitle

\section{Introduction}

In their seminal 1984 paper \cite{bb84}, Bennett and Brassard argued that some basic laws of physics may prove useful in cryptographic tasks. They considered first the task of \textit{key distribution} between distant partners and noticed that quantum signals are ideal trusted couriers: if the eavesdropper Eve tries to obtain some information, her action cannot remain concealed, because measurement modifies the state or, equivalently, because of the no-cloning theorem. In the second part of their paper, they turned to the task of bit-commitment and proposed a quantum solution relying on entanglement. In 1991, Ekert independently re-discovered quantum key distribution \cite{eke91}: his intuition was based on entanglement, more specifically on Bell's inequalities. These two works are the milestones of the field, even if precursors for these ideas had been brought up \cite{wiesner}.

The fact that security is based on physical laws lead to the hope that quantum cryptography may provide the highest possible level of security, namely security against an adversary with unrestricted computational power; in the jargon, \textit{unconditional security}. Further research vindicated only one of the two conjectures of Bennett and Brassard: key distribution can indeed be made unconditionally secure \cite{may96,lo99b,sho00}, while bit commitment cannot \cite{bitcom}. Most of the subsequent developments in quantum cryptography have therefore been devoted to \textit{quantum key distribution (QKD)}; several review papers are available \cite{gis02,dus06,sca08,lo08}.

\section{Quantum signals as incorruptible couriers}

Even before unconditional security was technically proved, \textit{``security based on the laws of physics''} became the selling slogan of QKD. It's catchy, and it can be understood correctly --- but it may also be understood wrongly and has often been explicitly spelled out as ``security based \textit{only} on the laws of physics''. Of course, a pause of reflection shows that the statement cannot possibly be as strong as that. For instance, the laws of physics do not prevent someone from reading the outcomes of a detector; however, if the adversary has access to that information, security is clearly compromised! But many people were just carried away by the power of the slogan --- fair enough, this does not happen only with QKD.

On the wings of enthusiasm, some promoters of QKD also managed to convey the impression that they were presenting \textit{the solution for (almost) every task of secret communication}. This may have impressed some sponsors. However, the main result was to alienate a great part of the community of experts in classical cryptography, who, unfamiliar with quantum physics though they may be, could not fail to spot the overstatement. Fortunately, several experts of QKD, well aware of the real scope of their research, managed to re-establish a constructive dialog. Both the interest and the niche character of QKD are generally admitted today.

In fact, the understanding of the niche character of QKD immediately clarifies the role of the laws of physics as well. The \textit{SECOQC White Paper} of 2007 \cite{all07} convincingly argued that \textit{QKD is a form of ``trusted courier''} i.e. a potential solution for those tasks, for which a trusted courier may be useful. For instance, if one can guarantee that a one-time pad key has not been revealed during its exchange, then the secret is guaranteed also in the future: this is an advantage over complexity-based schemes \cite{biham}. Now, with human couriers, we are fairly familiar. Suppose Alice creates a one-time pad key on her computer, burn it on a DVD and entrust to a human courier Charlie the task of bringing it to Bob. Alice should be confident that
\begin{itemize}
\item[(i)] her computer and Bob's are not leaking information, by themselves of through active hacking;
\item[(ii)] Charlie is honest at the moment of receiving the key from Alice;
\item[(iii)] During his travel from Alice to Bob, Charlie will neither be corrupted nor let information leak out inadvertently.
\end{itemize}
Replacing Charlie \textit{with quantum couriers, one does not have to worry about (iii) anymore: the laws of physics guarantee it; but they don't guarantee (i) and (ii)}. Indeed, it's pretty obvious that (i) must be enforced also for QKD. As for (ii), a ``dishonest'' quantum courier would be a quantum signal whose state has not been accurately characterized.

Still, one may think that the danger of (i) and (ii) does not extend beyond caricature examples: ``Sure enough, if Eve can see Alice through a window...; sure enough, if the source produces always two photons instead of one... But one can easily check for such blunders''. Unfortunately, exactly the opposite is the case: blunders affecting the security through failures of (i) or (ii) may be numerous and very subtle; most of the recent developments in practical QKD have to do with those concerns, as we are going to show in the next Section.

Before that, we want to stress an important, though quite obvious, issue: in this paper, we review the development of practical QKD. However, since assumptions (i) and (ii) are common for quantum and classical couriers, classical implementations must be checked as well for similar blunders. In other words, we are by no means implying that QKD would be more problematic than classical cryptography. In fact, QKD does have an advantage over classical methods, namely (iii).

\section{All that the laws of physics don't take care of}
\label{sec2}

\subsection{Problems at preparation}

We begin by examining the need for a careful assessment of the properties of the courier. Here is a list of examples. Note that most of them refer to implementations with weak coherent pulses: probably not because they are much worse than others, but because they have been scrutinized more thoroughly.

\begin{enumerate}
\item \textit{Problem:} attenuated laser pulses are not single photons, multi-photon components are important \cite{pns}. \textit{Solutions:} adapt the security proofs to take the effect into account \cite{gllp}, or change the protocol \cite{decoy,sarg} or of course change the source.

\item \textit{Problem:} successive pulses emitted by a laser are generally not independent, they may have phase coherence \cite{lopresk}. \textit{Solution:} adapt the security proofs (not done at the moment of writing) or actively randomize the phase.

\item \textit{Problem:} in the so-called ``plug-and-play'' implementations (the ones chosen for several commercial setups), photons do a round trip: Alice's device must receive light, code it and resend it \cite{muller}. But then, one must assume that the photons that enter Alice's lab might have been prepared by Eve \cite{gis06}. \textit{Solution:} add attenuation and active phase randomization, then use a suitable security proof \cite{zha08}.

\item \textit{Problem:} in continuous-variables QKD, if the local oscillator travels between Alice and Bob, the implementation is completely insecure unless Bob monitors the intensity \cite{hae08}. \textit{Solution:} add a beam-splitter and monitor the intensity.

\item \textit{Problem:} in some implementations, the different letters of the QKD alphabet are prepared by different light sources \cite{fourdiodes}. Each source may have its own fingerprint: for instance, even if coding is supposed to be in polarization, different sources may have different spectra. Also, minor initial or temperature-dependent differences in the electric driving circuitry of each source may go undetected in normal operation or assembly of the setup, but certainly leave a temporal fingerprint in the transmitted signal. \textit{Solution:} no miracle solution exists, one has to characterize the sources and bound the possible leakage of information.

\end{enumerate}

\subsection{Problems at detection}

Let us now review some examples of the problems at the level of detection. One such problem (admittedly, an anecdotal one) was stressed in the very first demonstration experiment presented by Bennett and coworkers \cite{benexp92}: the Pockels cells used to select the bases were driven by high-voltage devices, which made an audible sound depending on the basis or letter selection. Someone said that the device provided ``unconditional security against a deaf eavesdropper'': a joke... or a prophetic insight in the fate of practical QKD?

\begin{enumerate}
\item \textit{Problem:} an example of leakage of classical information explores parasitic properties of detectors. It is known that, upon detection, Silicon avalanche photodetectors emit light due to hot carrier recombination. This light may leak out through the optical channel, revealing which detector has fired~\cite{kurtsiefer:01a}. \textit{Solution:} other photo-diodes have been tested and no such back-flashing was detected (of course, these studies rely on the assumption that the devices used to probe for such radiation capture any sensibly accessible wavelength range).

\item \textit{Problem:} in ``plug-and-play'' systems, as mentioned, Alice's device is open to receive photons, before coding and resending them. The eavesdropper may implement a \textit{Trojan horse attack} to probe Alice's phase modulator: send in light (say at a different wavelength) and collect it back, coded. \textit{Solution:} because the setup involves attenuators, the additional light that is sent in should be quite intense; a proportional detector is then added at the entrance of the setup, which should detect unusually strong signals \cite{ribordy98}.

\item \textit{Problem:} light fields ('faked states') can be generated which force at least some of the common detectors to produce outcomes resembling those corresponding to the detection of single photons~\cite{makarov:05}. This may be exploited to implement something similar to a man-in-the-middle attack. \textit{Solution:} depends on the details of the implementation.

\item \textit{Problem:} photodetectors may also be manipulated to change their timing behaviour~\cite{makarov:06}, such that the detection time is partly correlated with the detection outcome. An experimental evaluation of this leakage channel became known as the time-shift attack~\cite{zhao08}. \textit{Solution:} check that all the detectors have the same timing statistics.

\item \textit{Problem:} in a similar fashion, communication of detection times (necessary in any scenario with a lossy communication channel) with a too high accuracy may reveal substantial information about the measurement results,  just due imbalanced electronic delays and/or detector parameter scatter \cite{lamas:07}. \textit{Solution:} do not reveal too many digits of your detection times.
\end{enumerate}

\subsection{A balance}

Just as in every field, there have been sheer mistakes in practical QKD: using an inadequate security bound (see below), neglecting to authenticate the classical channel, and the like. The problems reviewed above, however, do not belong to that category: each of them has been the object of a \textit{real discovery}.

The positive side of it all is that, once identified, each of those problems can be solved: thanks to these discoveries, the security of implementations has increased over the years. Another good piece of news is that, in spite of serious scrutinizing, there is no hint of a threat that would compromise the security of QKD in an irredeemable way. But there is a negative side to it: before being identified, each of the problems above represented a serious potential breach of security. It is a truism to stress that we may not be aware of similar problems, which have not been discovered yet.

We come to the bottom line of this section. We believe that this state of affairs cannot be simply dismissed with a ``there have been examples of \textit{bad design} of the device''. At any stage of development, the devices were actually carefully designed; the security claims of the authors were accepted as valid by referees and colleagues. Neither now, nor at any later time, will one be able to guarantee that the devices in use are provably good. And it is certainly not a good idea to close one's eyes, invoke the laws of physics and dump on them a responsibility they cannot possibly bear.

\subsection{On the use of security proofs}

We conclude this section with yet another series of concerns. Suppose for a moment that all the possible issues related to the implementation are under control. Can one finally rest in peace and trust the laws of physics? In principle one can, \textit{provided} all the assumptions, under which the security bounds were derived, are fulfilled by the implementation. Indeed, another dangerous shortcut consists in associating ``unconditional security'' with ``no assumptions'': no assumptions should be made on the power of the eavesdropper, but assumptions \textit{must} be made on what Alice and Bob are doing. Here are a few examples:

\begin{itemize}

\item Unconditional security bounds took some time to reach those who were evaluating or implementing practical schemes. The bounds that finally percolated were plagued by a huge assumption, of which very few people were aware: the bounds were valid only in the limit of infinitely long keys! \textit{Finite-key bounds} had already been addressed in some of the first security proofs \cite{may96,biham,inamori}; such bounds are now tractable and the whole community is aware of the stringency of finite-key corrections \cite{finitekey}.

\item Security proofs always imply a \textit{modeling of the detection process}. For instance, almost all the proofs in discrete-variable QKD are based on an assumption called \textit{squashing}: basically, whether a detector ends up squashing all the complexity of a state of the electromagnetic field (the ``real thing'') into a qubit (the thing theorists work with). The validity of this assumption was proved recently for the BB84 coding \cite{squashing}; for other protocols, it remains an assumption.

\item The most general theoretical bounds can accommodate all possible statistics. When applied to the study of a practical setup, however, simplifying assumptions are usually made. For instance, we are aware of only one proof, in which the detectors are allowed to have different efficiency, which is the case in reality \cite{fung}.

\end{itemize}

Ultimately, the suitable security bound for an implementation cannot be found explicitly spelled out in a theoretical paper; nor even in an experimental paper reporting on a similar implementation. Those papers should provide guidance, but each specific setup must be the object of a dedicated study --- whence another element of trust creeps in: one must trust the thoroughness of this dedicated study.

\section{Paths for the future}

QKD has evolved from the dreams of childhood to the seriousness of maturity. What is the next stage? Sure enough, ``only time will tell what we will do in the future'' \cite{constantin}. But the facts sketched above, combined with some tendencies within the QKD community, allow a guess of two directions in which the field may evolve in the coming years.

\subsection{Option 1: reasonable security of a device}

Although they do not provide ``security based only on the laws of physics'', usual QKD devices provide a quite reasonable level of security if implemented with technical competence and without false complacency on their ``quantumness''. After all, couriers of classical information must also be trusted; once the trust is there, QKD guarantees the incorruptibility of the courier during its travel --- a guarantee that classical information cannot offer.

Here we have therefore a \textit{first possible stance}: give up claims of ultimate security, find a competitive edge and try to produce devices than compare favorably with those operating with classical information. It is our impression that some of main actors in practical QKD have already taken this stance. Apart from technological development, one of the main challenges along the way will consist in finding where exactly the competitive edge may lie --- a superficial survey of claims may suggest that long-distance implementations are one of the goals, but in fact QKD does not seem to be a viable solution in that regime \cite{disadv}.

\subsection{Option 2: device-independent security and its price}

Recently, some authors have come up with a new class of QKD protocols that come as close as possible to the claim of ``security based only on the laws of physics''. The idea was already present in Ekert's 1991 seminal paper \cite{eke91}, but went unnoticed for many years. The key ingredient is that \textit{Bell's theorem is independent of quantum physics}. As a consequence, it is also independent of the details of the physical systems under study (its Hilbert space, its state, the measurements that are performed). Therefore, a protocol that estimates Eve's information through the amount of violation of a Bell-type inequality is ``device-independent'' \cite{devindep}.

Of course, even the security of device-independent protocols is not based only on the laws of physics; however, it seems that only those requirements are left that are \textit{strictly necessary}: the eavesdropper does not read your data, does not know which measurements you are going to choose, and the like. A comprehensive discussion can be found in Ref.~\cite{pironio}.

What we want to highlight here is that this level of security, arguably the most paranoid one can envisage, comes together with very stringent constraints. As an example, we show how \textit{device-independent protocols have an (almost) intrinsic limitation in distance}. This stems from the requirement that the detection efficiency must be high enough to close the so-called detection loophole. Indeed, as soon as the fraction of detected pairs falls below a given threshold, the observed violation of Bell-type inequality could have been created by pre-established agreement: the devices may just contain computers pre-programmed by the eavesdropper! In an implementation, being impossible to distinguish losses in transmission from losses due to the quantum efficiency of the detector, the threshold gives the value of the tolerable \textit{total} amount of losses.

How much is this threshold? A thorough study of the detection loophole is missing, partly because the classification of Bell's inequalities is a hard task in itself, and partly because the whole issue was considered of limited interest before the idea of device-independent QKD came about. It is known that there is no finite lower bound: there are explicit examples in which the detection loophole can be closed with arbitrarily low efficiency \cite{massar}. However, these examples use states of very large dimension $d$ and a number of measurements that is exponential in $d$. For a QKD protocol to qualify as ``practical'', the number of measurements and of outcomes must be kept ``reasonable''. The set of inequalities with few measurements and few outcomes has been studied in great detail (though not in its fullness) and, in those cases, the detection threshold is always found well above $50\%$ \cite{detloop}. For definiteness therefore, let us take the value of $50\%$ for the threshold of total detection efficiency \cite{lutk}.

InGaAs photodetectors in the telecom wavelength range have efficiencies below 30\% even under the most optimistic specifications; none of the implementation using those detectors can therefore be used for device-independent security. Depending on the trust into manufacturer specifications, setups using Silicon-APD may just about reach the threshold on the detector side, but then almost no losses can be tolerated in the channel. Assuming detectors with 100\% efficiency \cite{seawoonam}, fiber-optical transmission channels without any interconnect losses and with the usually quoted (optimistic) attenuation coefficient of 0.18\,dB/km would limit a direct QKD link to a distance of 16\,km. 

This issue of the distance has been presented as an example of how stringent the requirements for device-independent security may be. If we believe that history repeats itself, further scrutiny will lead to identifying further limitations of these very recent protocols. In other words, it may be premature to attach any practical value to device-independent QKD; whence the \textit{second possible stance} that we envisage for the coming years: focus on developing the tools (both theoretical and experimental) required to demonstrate the ultimate level of security, leaving aside, at least temporarily, all claims of usefulness.

\section{Conclusion}

We have reviewed the evidence of the fact that QKD guarantees security based on the laws of physics \textit{provided} the implementation is perfect, or more precisely, provided all the imperfections of the implementation have been characterized and their effect is accounted for. Since a thorough check of all possible leakage channels is impossible, the security of any specific implementation of QKD will always be based on some elements of trust. Expressions like ``security based only on the laws of physics'' or ``unconditional security'' are convenient among experts, as part of their technical jargon; but when they leak out to larger audiences, they almost invariably convey the wrong message (the same is true for ``device-independent'', though fortunately this expression has not reached the general public yet).

Specifically, we argued that the level of trust for usual QKD protocols, like BB84, is not very different from the one demanded of a ``trusted courier'' carrying classical information. Protocols based on Bell's inequalities minimize the number of elements of trust but come with very stringent requirements. 

Twenty-five years after BB84, the field of QKD seems on the point of splitting in two directions: (i) the development of prototypes optimized for the needs of niche tasks and guaranteeing a ``reasonable'' level of security; (ii) the quest for demonstrating the most paranoid level of security, leaving aside, at least temporarily, the claim of practical usefulness.

\textbf{Note added in proof:} this text reflects the state of practical QKD in 2009, i.e. on the 25th anniversary of the BB84 protocol. Among the progresses in the last two years, two are worth at least mentioning. First, complete hacking of QKD systems has been reported, exploiting a loophole in the hardware, which can be closed \cite{makarov,ilja}. Second, methods have been suggested that would avoid the distance limitation in device-independent QKD \cite{gps,cm}. While it is too early to assess the practical value of these methods, it is encouraging to see that, when the real challenges are clearly spelled out, the inventiveness of physicists rises to meet them.

\section*{Acknowledgments}

This work is supported by the National Research Foundation and Ministry of Education, Singapore. We thank Antonio Ac\'{\i}n, Alexios Beveratos, Nicolas Gisin, Antia Lamas-Linares, Vadim Makarov, Gr\'egoire Ribordy and Hugo Zbinden for illuminating exchanges on these topics over the past few years; and the editors Renato Renner and Tal Mor for their support and their clarifying comments.

\end{document}